\documentclass[aps,superscriptaddress,preprintnumbers,showpacs]{revtex4}
\usepackage{amssymb}
\usepackage{dcolumn}
\usepackage{bm}
\usepackage{amsmath}
\usepackage{mathrsfs}
\usepackage{slashed}
\usepackage{graphicx}
\begin{document}
\bibliographystyle{plain}

\preprint{arXiv:1111.7050v4 [hep-ph]}

\newcommand*{\PKU}{School of Physics and State Key Laboratory of Nuclear Physics and
Technology, Peking University, Beijing 100871, China}\affiliation{\PKU}
\newcommand*{\chep}{Center for High Energy Physics, Peking University, Beijing 100871, China}\affiliation{\chep}
\newcommand*{\chps}{Center for History and Philosophy of Science, Peking
University, Beijing 100871, China}\affiliation{\chps}

\title{ The Phantom of the OPERA: Superluminal Neutrinos\footnote{This report is based on a number of talks and seminars:
talks at CosPA2011, October 28-31, 2011, Peking Univ., Beijing, at
Workshop on LHC and Dark Matter Physics, November 24-26, 2011,
Shanghai JiaoTong Univ., Shanghai, and at Workshop on Frontiers in
Theoretical Physics by State Key Laboratory of Theoretical Physics,
December 26-28, 2011, ITP@CAS, Beijing; seminars at Taiwan Univ. at
Taipei, Central Univ. at Chungli, Tsing Hua Univ. at Hsinchu, and
Peking Univ. at Beijing. This reviewed is published as B.-Q.Ma,
Mod.Phys.Lett.A 27, 1230005 (2012).}}

\author{Bo-Qiang Ma}\email{mabq@pku.edu.cn}\affiliation{\PKU}\affiliation{\chep}\affiliation{\chps}

\begin{abstract}
This report presents a brief review on the experimental measurements
of the muon neutrino velocities from the OPERA, Fermilab and MINOS
experiments and that of the (anti)-electron neutrino velocities from
the supernova SN1987A, and consequently on the theoretical attempts
to attribute the data as signals for superluminality of neutrinos.
Different scenarios on how to understand and treat the background
fields in the effective field theory frameworks are pointed out.
Challenges on interpreting the OPERA result as a signal of neutrino
superluminality are briefly reviewed and discussed. It is also
pointed out that a covariant picture of Lorentz violation may avoid
the refutation on the OPERA experiment.
\end{abstract}


\maketitle

\section{The Phantom of the OPERA}

The report that the OPERA Collaboration has unveiled evidence for a
faster-than-light speed of muon neutrinos has caused a stir among
physical society as well as public media. The light speed $c$ is
considered as the uppermost high speed of any kind of particles in
special relativity, thus the OPERA experiment puts up a strong
challenge to Einstein's theory of relativity. Since the release of
the OPERA paper~\cite{Adam2011} on September 23 of 2011, there have
been a fast growing number (over a hundred) of papers in the arXiv,
discussing the the OPERA neutrino anomaly from various aspects. The
OPERA anomaly of neutrinos is just like ``the Phantom of the Opera"
behind the mask, we are still unclear whether it is a ghost or
something else at this stage.

OPERA stands for ``the Oscillation Project with Emulsion-tRacking
Apparatus'', which is an instrument for the investigation on
neutrino oscillations. The experiment~\cite{Adam2011} is a
collaboration between CERN in Geneva, Switzerland, and the
Laboratori Nazionali del Gran Sasso (LNGS) in Gran Sasso, Italy. It
exploits neutrinos from CERN to Gran Sasso (CNGS). The muon
neutrinos in the CERN-CNGS neutrino beam were detected by the OPERA
detector over a baseline of about 730~km. Compared to the time taken
for neutrinos traveling at the speed of light in vacuum, an earlier
arrival time of $(60.7\pm6.9~(\mathrm{stat.})\pm
7.4~(\mathrm{sys.}))~{\mathrm{ns}}$ was measured. The neutrino
velocity $v$ is thus measured and its difference with respect to the
vacuum light speed $c$ is ${(v-c)}/{c}=\left(2.48 \pm 0.28
(\mathrm{stat.}) \pm 0.30 (\mathrm{sys.})\right)\times 10^{-5}$ at a
significance of $6\sigma$.

The OPERA result surprised people because of its magnitude of the
superluminality of $10^{-5}$, which is bigger compared to any
previous constraints. Another reason is because of its high
precision with big significance, which means that the conclusion
should be reliable to an undoubtable confidence level if one can not
find error in the experiment. In fact, there have been similar long
baseline experiments of the same kind. The first direct measurement
of neutrino velocity has been performed at Fermilab thirty years
ago~\cite{Alspector1976,Kalbfleisch1979}. Based on 9,800 events,
measurements of the velocity of muon neutrinos with energy ranging
from $30$~GeV to $200$~GeV gave
$|\beta_{\nu}-1|<10^{-5}$,
where $\beta_{\nu}\equiv v_{\nu}/c$. Just a few years ago, by using
the NuMI neutrino beam, the MINOS Collaboration~\cite{Adamson2007}
analyzed a total of 473 far detector neutrino events with an average
energy $~3 $~GeV. They reported a shift with respect to the expected
time of flight of
$\delta_t=-126\pm32~(\mathrm{stat})\pm64~(\mathrm{sys)}~\mathrm{ns}$,
which corresponds to a constraint on the muon neutrino velocity,
$(v_{\nu}-c)/c=(5.1\pm2.9)\times10^{-5}$
at $68\%$ confidence level. This $1.8\sigma$ signal was considered
to be compatible with also zero, therefore it does not provide a
strong evidence of Lorentz violation effects. However, with the new
measurement of OPERA detector, it is surprising to notice that the
Fermilab and MINOS results are compatible with the OPERA result,
though at much lower statistics.

Besides the long baseline experiments, there are also measurable
phenomenologies of superluminal neutrinos in astrophysics. For
instance, supernova explosion (SNe) is an extremely luminous event,
which causes a burst of radiation that outshines an entire galaxy.
The radiation includes photons in a board range of spectrum, as well
as neutrinos. Actually, most energy of a SNe is released in the form
of neutrinos, however, due to the weak interactions of neutrinos
with matters, only one event was observed with neutrino emissions on
23 February 1987, 7:35:35 UT ($\pm1$ min)
--- the Supernova 1987A in the Large Magellanic
Cloud~\cite{Hirata1987,Bionta1987}, which was optically observed on
24 February 1987. More than ten neutrinos were recorded with a
directional coincidence within the location of supernova explosion,
several hours before the optical lights were observed. Because of
weak interactions, neutrinos leak out of the dense environment
produced by the stellar collapse before the optical depth of photons
becomes visible. Hence an early-arrival of neutrinos is expected.
The journey of propagation of photons and neutrinos are of
astrophysical distance ($\sim 51.4$~kpc), hence it provides a unique
opportunity to measure~\cite{Longo1987} the speed of neutrinos to be
within the light speed with a precision of $\sim2\times10^{-9}$.

Interestingly, while the OPERA results seem to be in remarkable
consistence with other terrestrial muon neutrino velocity
measurements, they contradict with the Supernova 1987A neutrino
observation severely. The neutrinos detected from the Supernova are
mainly anti-electron neutrinos with energy around 10~MeV, rather
than 10~GeV scale muon neutrinos in the Fermilab, MINOS, and OPERA
experiments. The electron neutrinos of SN1987a also put bound on the
deviation of the velocity of neutrinos $v_{\bar{\nu}_e}$ with
respect to the light speed $c$: $|(v_{\bar{\nu}_e}-c)/c|\leq 2\times
10^{-9}$. The experimental data or bounds on the neutrino velocities
are listed in Table~\ref{data}. As the observation of neutrinos from
the Supernova is more reliable in time measurement, people also take
this confliction as a severe challenge to the correctness of the
OPERA experiment. It is expected that detection of neutrinos from
future galactic supernova will be helpful not only for the study of
neutrino velocities~\cite{Ellis:2008fc}, but also for the
investigation of other neutrino properties~\cite{Huang:2010qh}. The
detection of cosmogenic neutrino spectrum is also useful to learn
about the Lorentz violation effects in the neutrino
sector~\cite{Mattingly:2009jf}.

\begin{table}[h]
\caption{ Data/bounds on neutrino velocities from OPERA,
  Fermilab, MINOS, and
  SN~1987A data}
{\begin{tabular}{@{}lccccccccc@{}}  \toprule OPERA & Energy~(GeV) &
~~13.9~~ & ~~42.9~~ & & & & & &
\\ & $\frac{v_{\nu_\mu}-c}{c}$ ($10^{-5}$) & $2.17\pm0.83$ & $2.74\pm0.80$ & & &
& & & \\
\colrule Fermilab & Energy~(GeV) & ~32~ & ~44~ & ~59~ & ~69~ & ~90~
& ~120~ & ~170~ & ~195~ \\ & $\frac{v_{\nu_\mu}-c}{c}$ ($10^{-5}$) &
$-2^{+2}_{-3}$ & $2\pm7$ & $-1^{+2}_{-3}$ & $-1^{+2}_{-3}$ &
$1^{+3}_{-4}$ & $1\pm7$ & $1^{+2}_{-3}$ & $6^{+3}_{-4}$ \\ \colrule
MINOS & Energy~(GeV) & ~~3~~ & & & & & & &
\\ & $\frac{v_{\nu_\mu}-c}{c}$ ($10^{-5}$) & $5.1\pm{2.9}$ & & & & & & &
\\  \colrule SN1987A & Energy~(MeV) & $\sim~10$~~ & & & & &
& &
\\ & $\frac{v_{\bar{\nu}_e}-c}{c}$ ($10^{-9}$) & $\leq{2}$ & & & &
& & & \\ \botrule
\end{tabular}\label{data}}
\end{table}

After the first release of the OPERA result, there have been many
criticisms and doubts on the correctness of the experiment, such as
whether the clocks at the two sides of CERN and LGNS are correctly
adjusted by GPS technique as well as whether the distance between
the two sides is properly measured, and whether the beam duration
treatment of the data can introduce bias in the neutrino arrival
time measurement. The OPERA collaboration repeated~\cite{re-OPERA}
the measurement over the same baseline without any assumptions about
the details of neutrino production during the spill, such as energy
distribution or production rate, by using a new CERN beam which
provided proton pulses of 3~nanoseconds each with 524~nanosecond
gaps. Without using the earlier statistical computation, the OPERA
collaboration measured twenty events indicating neutrinos had
traveled faster than light by 60~ns, with 10~ns uncertainty. The
error bounds for the original superluminal speed fraction were
tightened further to $(2.37 \pm 0.32 (\mathrm{stat.}) + 0.34/ - 0.24
(\mathrm{sys.}))¡Á10^{-5}$, with the new significance level becoming
$6.2\sigma$.

The neutrino speed anomaly of the OPERA is now a ``phantom". The
re-confirmation of the OPERA result reminds us that the game just
begins, it is far from the end. Whether we believe it or not, the
phantom of the OPERA is there, deep down below the earth with a
mask. We need to finger it out, whether it is a ghost or an angel of
music.

\section{The OPERA Phantom as a Signal for Lorentz Violation}

Nowadays, there has been an increasing interest in Lorentz
invariance Violation (LV or LIV) both theoretically and
experimentally~\cite{ShaoMa10}. Special relativity, or Lorentz
symmetry, is one of the foundations of modern physics and has been
proved to be valid at very high precision. However, the possible
Lorentz symmetry violation (LV) effects are sought for decades from
various theories, motivated by the unknown underlying theory of
quantum gravity together with various phenomenological
applications~\cite{Kostelecky2011,lv3,Shao2010,Shao2011,lv4,Bietenholz:2008ni}.
This can happen in many alternative theories, e.g., the doubly
special relativity
(DSR)~\cite{Amelino-Camelia2002,Magueijo:2001cr,Zhang2011}, torsion
in general relativity~\cite{LV-GR1,Ni:2009fg,LV-GR2}, non-covariant
field
theories~\cite{Copenhagen1,Copenhagen2,Copenhagen3,Copenhagen4}, and
large extra-dimensions~\cite{Ammosov2000,Pas2005}. From basic
consideration, there were investigations on the concepts of
space-time such as whether the space-time is discrete or
continues~\cite{p99,Snyder,xu-l}, or whether a fundamental length
scale should be introduced to replace the Newtonian constant
$G$~\cite{lv5}. It has been revealed from physical arguments that
space-time is discrete rather than continuous~\cite{xu-l}, and we
also know that the introduction of the minimal length scale can be
manifested through the Lorentz violation~\cite{lv5}. The existence
of an ``{\ae}ther" or ``vacuum" can also bring the breaking down of
Lorentz invariance~\cite{Dirac,Bjorken}.

Although the OPERA result has been largely debated, neutrino
velocity anomaly appears to be a strong challenge to the well-known
fact in special relativity that no physical particle travels faster
than the light. One of the fundamental principles of relativity is
the Lorentz invariance, which states that the equations describing
the laws of physics have the same form in all admissible frames of
reference. Therefore it seems a natural attempt to attribute the
OPERA anomaly as a signal for Lorentz invariance violation. The
first a few papers~\cite{Camelia11,Cacciapaglia11} on the OPERA
anomaly are phenomenological analysis to use some kind of modified
dispersion relations to fit the data. Consequently there are a
number of
papers~\cite{SMS-OPERA,Qin-OPERA,SMS-OPERA2,Pfeifer:2011ve,Li:2011zm,Saridakis:2011eq,Nojiri:2011ju,Xue:2011tz,Chang:2011td,Yan:2011jm,Brodsky:2011nc}
to seek for the possibilities for superluminal neutrinos from
Lorentz violation in various theoretical models. In fact, the
possibilities of superluminal neutrinos were proposed with an
earlier version of standard model extension (SME)~\cite{Chodos85}
and with extra-dimensions~\cite{Ammosov2000,Pas2005} before the
OPERA experiment. There were also
attempts~\cite{Xiao08,Yang09,Liu:2011nwa,Diaz:2010ft} to reproduce
the neutrino oscillations through Lorentz violation rather than from
mass difference as in the conventional treatment~\cite{Ma:2011gh}.

Here I focus on the attempts~\cite{SMS-OPERA,Qin-OPERA,SMS-OPERA2}
to attribute the OPERA anomaly as a signal of Lorentz violation in
the effective field theory frameworks based on traditional
techniques in particle physics.  A useful model on Lorentz violation
is the minimal Standard Model Extension (SME), in which Lorentz
violation terms are constructed with standard model fields and
controlling coefficients are added to the usual standard model (SM)
Lagrangian~\cite{Colladay1998}. The origins for such LV operators
are suggested in many ways, of which spontaneous Lorentz symmetry
breaking proposed first in string theory is widely
recognized~\cite{Kostelecky1989}. The minimal SME is first applied
in Ref.~\cite{Qin-OPERA} to confront with the OPERA result as an
indication for superluminal neutrinos. There is also a recent
proposal to derive some supplementary LV terms from standard model
with a basic principle of the physical invariance with respective to
the mathematical background manifolds~\cite{Ma10,SMS3}, and such a
standard model supplement (SMS) framework has been applied to
discuss the Lorentz violation effects for the cases of Dirac
particles~\cite{Ma10}, photons~\cite{SMS3,Ma10graal,SMS-photon2},
and neutrinos~\cite{SMS-OPERA}, in which the superluminal neutrinos
as a signal of Lorentz violation was suggested. Here we do not go
into theoretical details, but outline the basic concepts why the
effective Lagrangians of effective field theory frameworks can bring
the superluminality of neutrinos and how one can handle the Lorentz
violation effects in such frameworks.

The general effective field theory framework starts from the
Lagrangian of the standard model, and then includes additional terms
containing the Lorentz violation effects. The magnitudes of these LV
terms can be constrained by various experiments. The Coleman-Glashow
model~\cite{Coleman99} is a most simple version of the effective
field theory framework with a scaler constant as the LV parameter.
In the minimal version of the SME~\cite{Colladay1998}, the LV terms
are measured with several tensor fields as coupling constants, and
modern experiments have built severe constraints on the relevant
Lorentz violation parameters~\cite{Kostelecky2011}. In the SMS
framework~\cite{Ma10,SMS3}, the LV terms are brought about from a
basic principle denoted as the physical independence (or physical
invariance), which requires that the equations describing the laws
of physics have the same form in all admissible mathematical
manifolds. Such principle leads to the introduction of some Lorentz
violation matrices $\Delta^{\alpha \beta}$ to the standard model
particles under consideration, with the elements of these LV
matrices to be measured or constrained from experimental
observations rather than from theory at first. Therefore one may
also consider the LV matrices in the SMS framework as similar to the
background tensor fields in the minimal SME model.

The Coleman-Glashow model~\cite{Coleman99} is a simple version to
include Lorentz violating terms into the standard model Lagrangian.
Let $\Psi$ denote a set of $n$ complex scalar fields assembled into
a column vector. With the invariance under the $U(1)$ group
$\Psi\rightarrow e^{-i\lambda}\Psi$, the most general free
Lagrangian is: $ {\cal L}
=\partial_\mu\Psi^*Z\partial^\mu\Psi-\Psi^*M^2\Psi$, where $Z$ and
$M^2$ are positive Hermitian matrices. One can always linearly
transform the fields to make $Z$ the identity and $M^2$ diagonal,
thus obtaining the standard theory of $n$ decoupled free fields. One
can then add to the standard model Lagrangian the Lorentz-violating
term:
\begin{equation}
{\cal L}\rightarrow{\cal L} +
\partial_i\Psi \epsilon\partial^i\Psi,
\end{equation}
where $\epsilon$ is a Hermitian matrix that signals the Lorentz
violation in the Coleman-Glashow model.

The SME Lagrangian in the neutrino sector takes the
form~\cite{Colladay1998,Yang09,Qin-OPERA}
\begin{align}
\mathcal{L}=\frac{1}{2}i\overline{\nu}_{A}\gamma^{\mu}\overleftrightarrow{D_{\mu}}\nu_{_B}\delta_{_{AB}}
+\frac{1}{2}ic_{_{AB}}^{\mu\nu}\overline{\nu}_{_A}\gamma^{\mu}\overleftrightarrow{D^{\nu}}\nu_{_B}
-a_{_{AB}}^{\mu}\overline{\nu}_{_A}\gamma^{\mu}\nu_{_B}+\cdots\;,\label{Lagrangian}
\end{align}
where $c_{AB}^{\mu\nu}$ and $a_{AB}^{\mu\nu}$ are Lorentz violation
coefficients resulting from tensor vacuum expectation values in the
underlying theory, the subscripts $A,B$ are flavor indices, and the
ellipsis denotes the non-renormalizable operators (eliminated in the
minimal SME). The first term in Eq.~(\ref{Lagrangian}) is exactly
the SM operator, and the second and third terms (CPT-even and
CPT-odd respectively) describe the contribution from Lorentz
violation.

For the electroweak interaction sector, the Lagrangian of fermions
in the SMS framework can be written as~\cite{Ma10,SMS3,SMS-OPERA}
\begin{eqnarray}\label{Lagrangian_F}
\mathcal{L}_{\mathrm{F}} &=&
i\bar{\psi}_{A,\mathrm{L}}\gamma^{\alpha}\partial_{\alpha}\psi_{B,\mathrm{L}}\delta_{AB}+
i\Delta^{\alpha\beta}_{\mathrm{L},AB}\bar{\psi}_{A,\mathrm{L}}\gamma_{\alpha}\partial_{\beta}\psi_{B,\mathrm{L}}\nonumber\\
&&
+i\bar{\psi}_{A,\mathrm{R}}\gamma^{\alpha}\partial_{\alpha}\psi_{B,\mathrm{R}}\delta_{AB}+
i\Delta^{\alpha\beta}_{\mathrm{R},AB}\bar{\psi}_{A,\mathrm{R}}\gamma_{\alpha}\partial_{\beta}\psi_{B,\mathrm{R}},
~~~~
\end{eqnarray}
where $A,B$ are flavor indices. The Lorentz violation terms are
uniquely and consistently determined from the standard model by
including the Lorentz violation matrices $\Delta^{\alpha\beta}$,
which are generally particle-dependent~\cite{SMS3} with flavor
indices. For leptons, $\psi_{A,\mathrm{L}}$ is a weak isodoublet,
and $\psi_{A,\mathrm{R}}$ is a weak isosinglet. After the
calculation of the doublets and classification of the Lagrangian
terms again, the Lagrangian can be written in a form like that of
Eq.~(\ref{Lagrangian_F}) too. Assume that the Lorentz violation
matrix $\Delta^{\alpha\beta}_{AB}$ is the same for the
left-handedness and right-handedness, namely
$\Delta^{\alpha\beta}_{\mathrm{L},AB}=\Delta^{\alpha\beta}_{\mathrm{R},AB}=\Delta^{\alpha\beta}_{AB}$.
Without considering mixing between flavors, one can rewrite
Eq.~(\ref{Lagrangian_F}) as
\begin{equation}\label{Lagrangian_F_total}
\mathcal{L}_{\mathrm{F}}=\bar{\psi}_A
(i\gamma^{\alpha}\partial_{\alpha}-m_A)\psi_A
+i\Delta^{\alpha\beta}_{AA}\bar{\psi}_A\gamma_{\alpha}\partial_{\beta}\psi_A,
\end{equation}
where $\psi_A=\psi_{A,\mathrm{L}}+\psi_{A,\mathrm{R}}$, i.e., the
field $\psi_A$ is the total effects of left-handed and right-handed
fermions of the given flavor $A$.  When there is only one handedness
for fermions, $\psi_A$ is just the contributions of this one
handedness, which is the situation for neutrinos. The mass term in
the Lagrangian $\mathcal{L}_{\mathrm{F}}$ is included, one can let
$m_A\rightarrow 0$ for massless fermions.

However, as the LV terms in the general effective field theory
frameworks can be considered as added by hands rather than from
basic theories, there exist different scenarios on how to understand
these background fields and also on how to handle them. We list
three options of possible understandings and treatments:

\begin{itemize}
\item
{\bf Scenario I}: which can be called as fixed background scenario
in which
the backgrounds are taken as fixed parameters in any inertial frame
of reference the observer is working. It means that the backgrounds
can be taken as with the same formalism and with approximately the
same parameters for any working reference frames such as earth-rest
frame, sun-rest frame, or CMB frame for an observer. This scenario
can be adopted when the observer is focusing on the Lorentz
violation effect within a certain frame and does not care about
relations between different frames, or the situation could become
very complicated with different formalisms in different frames from
the requirement of consistency. This scenario can apply as a
practical tool for all of the above mentioned three versions of the
effective field theory framework: the Coleman-Glashow model, the
minimal SME, and the SMS framework.


\item
{\bf Scenario II}: which can be called as ``new {\ae}ther" scenario
in which the background fields transform as tensors between
different inertial frames of reference but keep unchanged within the
same frame. It means that there exists a privileged inertial frame
of reference in which the background can be considered as the ``new
{\ae}ther", i.e., the ``vacuum" at rest, which changes from one
frame to another frame by Lorentz transformation. Within a same
frame of reference, these background fields are just treated as
fixed parameters. This scenario cannot apply directly to the
Coleman-Glashow model, as in this model the LV parameter is a scaler
which should keep invariant in any working reference frames, but it
can apply to the minimal SME and also to the SMS. In fact, the
previous phenomenalogical analysis in the minimal SME are based on
this scenario.

\item
{\bf Scenario III}: which can be called as covariant scenario in
which the background fields transform as tensors adhered with the
corresponding standard model particles. It means that these
background fields are emergent and covariant with their standard
model particles. This scenario cannot apply to the Coleman-Glashow
model, but can apply to the minimal SME and also to the SMS. Such a
scenario probably still has not been considered in previous studies
in the effective field theory with normal Lorentz transformation.
\end{itemize}

Now we come to the question how the Lorentz violation effects can
exist in the different scenarios of Lorentz violation in the
effective field theory framework.
Generally, the mass energy relation of particles with 4-momentum
$p_{\mu}$ and mass $m$ can become
\begin{equation}\label{mass_energy_LV_fermion}
p^2=m^2+\lambda(\Delta_1,\cdots,\Delta_j,\cdots,\Delta_n,p),\quad
1\leq j\leq n
\end{equation}
where the parameters/tensors $\Delta_j$ represent background fields
which influence the corresponding standard model particles and
$\lambda$ is a new quantity that signals the Lorentz invariance
violation. Different frameworks of Lorentz violation have different
parameters $\Delta_j$. Within the standard model, we cannot observe
these background fields. So we just focus on the particles of the
standard model, make Lorentz transformations on these particles, and
then find that all equations are Lorentz invariant. But when we add
terms including $\Delta_j$ coupling with these particles and still
make Lorentz transformation on just these same particles, the
equations with the added terms are found to violate Lorentz
invariance this time. So in this sense, the added terms violate
Lorentz invariance. Thus $\Delta_j$ are the tensors violating
Lorentz invariance. On the other hand, when the background tensors
$\Delta_j$ are Lorentz covariant with these particles too, the
equations are still Lorentz invariant, because all the space-time
indices in the added terms are contracted generally. That is to say,
when both $p$ and $\Delta_j$ of Eq.~(\ref{mass_energy_LV_fermion})
are Lorentz transformed to $\hat{R}(p)$ and $\hat{R}(\Delta_j)$
respectively, Eq.~(\ref{mass_energy_LV_fermion}) is still Lorentz
invariant. So the Lorentz violation frameworks (e.g. the SMS and the
SME) violate the common Lorentz invariance of the standard model,
but keep the Lorentz invariance in the new sense that the background
Lorentz violation tensors are Lorentz covariant too. With the
terminology for Lorentz violation theories, we say that the Lorentz
violation frameworks break the Lorentz invariance but keep the
Lorentz covariance. It is also possible that the added terms break
both the Lorentz invariance and the Lorentz covariance, such as the
situation of the Coleman-Glashow model, but in this case the
introduction of different formalisms between different reference
frames can not be avoided for consistency and this makes things very
complicated. In all of the above three scenarios of Lorentz
violation, the motions of these standard model particles are
influenced by the background fields, therefore the Lorentz violation
effects do exist as compared with the situation without background
fields. Then the energy-momentum dispersion relation of a particle
is different from that in the free case. One thus can calculate the
particle velocity through the new dispersion relation, in which the
background fields enter as parameters. The velocity of a particle
could be therefore superluminal or subluminal by adjusting the LV
parameters. By confronting with the OPERA result, the LV parameters
are estimated in Ref.~\cite{SMS-OPERA} for the SMS framework and in
Ref.~\cite{Qin-OPERA} for the minimal SME.


As there are a large number of parameters in the minimal SME and the
SMS framework, there are still large degrees of freedom to fit the
OPERA, Fermilab, MINOS and supernova data for superluminal
neutrinos. Therefore it is still too early to suggest a specific
model at the moment but we refer to Refs.~\cite{SMS-OPERA,Qin-OPERA}
for some possible choices of simple toy models to confront with
data. However, we noticed that to reconcile the difference between
the muon neutrino data of OPERA, Fermilan and MINOS and the electron
neutrino data of supernova, we may investigate along two possible
directions:

\begin{itemize}
\item
The flavor dependence: the observation that the species of supernova
neutrinos are different from those of terrestrial neutrinos --- the
former being electron (and/or anti-electron) neutrinos, while the
measured collider neutrinos from OPERA are muon neutrinos. We
suspect a family hierarchy should be responsible for the observed
different velocities~\cite{SMS-OPERA,flavor-difference}. In the
dispersion relations, Lorentz violation coefficients of different
flavors are generally different, hence if there exist family
hierarchies of these parameters, the different propagation behaviors
of supernova neutrinos and terrestrial muon neutrinos can be
understood.

\item
The energy dependence: the observation that the supernova electron
neutrinos are of 10~MeV scale while the collider muon neutrinos are
of 10~GeV scale may lead to possible energy dependence in the
dispersion relation to reconcile with the different
superluminalities for collider neutrinos and supernova neutrinos.
There have been a number of models~\cite{Chang:2011td,Zhao:2011sb}
can realize such a requirement phenomenalogically.
\end{itemize}

There are also other ideas on the different neutrino velocities
between OPERA and supernova, such as by including the matter effect
in the earth crust~\cite{Ciuffoli:2011ji} and by the introduction of
sterile neutrinos which may take a shortcut in
propagation~\cite{Hannestad:2011bj,Nicolaidis:2011eq,Marfatia:2011bw}.

\section{A Challenge on the Rationality of the OPERA Result}

Cohen and Glashow argued~\cite{Glashow11} that if the Lorentz
violation of the OPERA experiment is of $10^{-5}$, the high energy
muon neutrinos exceeding tens of GeVs can not be detected by the
Gran Sasso detector, mainly because of the energy-losing process
$\nu_\mu\rightarrow\nu_\mu+e^{+}+e^{-}$ analogous to Cherenkov
radiations through the long baseline about 730~km. Bi {\it et al.}
also argued that the Lorentz violation of muon neutrinos of order
$10^{-5}$ will forbid kinematically the production process of muon
neutrinos $\pi\rightarrow \mu + \nu_\mu$ for muon neutrinos with
energy larger than about 5~GeV~\cite{Bi11}. Such arguments put up a
strong challenge to the rationality of the OPERA experiment and the
consequent suggestion to attribute the OPERA experiment as a signal
of Lorentz violation.

As we have pointed out, the Cohen-Glashow argument is based on the
Coleman-Glashow model~\cite{Coleman99} of Lorentz violation, and
their argument is based on some implicit assumptions, such as that
the Lorentz violation is measured by just a scalar Lorentz violation
parameter. Such a conclusion is not valid in general in other
Lorentz invariance violation frameworks. A response to Cohen,
Glashow  is offered in Ref.~\cite{Smolin11}, where Amelino-Camelia,
Freidel, Kowalski-Glikman and Smolin argued that the energy
threshold for the anomalously Cherenkov analogous process
$\nu_\mu\rightarrow\nu_\mu+e^{+}+e^{-}$ makes physical event
observer-dependent. They pointed out that the deformed Lorentz
transformation can avoid the problem brought about by these
arguments. There have been a number of
investigations~\cite{Li:2011ad,Ling:2011re,Huo:2011ra,Guo:2011uq},
indicating that these analogous Cherenkov radiations can be avoided
by adopting some forms of deformed Lorentz transformations. The
frameworks with deformed Lorentz transformation, such as the doubly
special
relativity~\cite{Amelino-Camelia2002,Magueijo:2001cr,Zhang2011},
might be cataloged into the covariant picture of Lorentz violation.

It is also pointed in Ref.~\cite{SMS-OPERA2} that the Cohen-Glashow
argument is not valid in general in other Lorentz invariance
violation frameworks in which the Lorentz invariance is breaking
whereas the Lorentz covariance still holds, such as the standard
model supplement (SMS)~\cite{Ma10,SMS3} or the standard model
extension (SME)~\cite{Colladay1998}. The derived dispersion
relations in the minimal SME and the SMS might be treated with an
option as covariant with the momentum of the muon neutrino and thus
can avoid the Cherenkov-like radiations.

It has been reported~\cite{ICARUS} by the ICARUS Collaboration that
there is no evidence for the analogues Cherenkov radiation of muon
neutrinos from CERN to the LNGS, where the OPERA experiment is also
performed. If taking the arguments of Cohen-Glashow and Bi {\it et
al.} as true, then one must refute the OPERA result of neutrino
superluminality as Ref.~\cite{ICARUS} did. However, we take the
ICARUS result as a support of our argument on the forbidding of
these Cherenkov-like processes, rather than a refutation of the
OPERA result. Therefore our argument can accommodate both the OPERA
and the ICARUS experiments, whereas one must refute the OPERA result
of superluminality or the ICARUS result of no analogues Cherenkov
radiation based on the arguments for these Cherenkov-like processes.
It is not adequate to refute an experimental observation by just
pure theoretical argument, but instead, reliable experimental
observations can be used to rule out theoretical arguments. From a
phenomenalogical viewpoint, what reported by the ICARUS
Collaboration of no Cherenkov-like radiation is just among one part
of the observation by the OPERA Collaboration already. What they did
is just to rely on a theoretical argument to refute the
superluminality part of the OPERA result.

The covariant picture of Lorentz violation, whether in the effective
field theory or in some kind of frameworks with deformed Lorentz
transformation, might accommodate the superluminality of neutrinos
with no Cherenkov like radiation. This might provide a possible way
for a consistent approach to handle the Lorentz violation effects.

\section{Conclusion}

Researches on Lorentz violation have been active for many years,
with various theories have been proposed and many phenomenological
studies have been performed to confront with various observations.
Though there have been many phenomena which could be marginally
considered as possible evidences or hints for Lorentz violation,
there is no convincing evidence yet, including the OPERA anomaly,
which is still a phantom under a mask. There are still many
challenges to attribute the OPERA anomaly as an evidence for Lorentz
violation, however, the OPERA anomaly provides a new chance for
Lorentz violation study. The re-confirmation of the experiment by
the OPERA collaboration itself reminds us that ``the phantom of the
OPERA" just begins, and we still need some stages to reveal the true
face of this OPERA phantom. We conclude that Lorentz violation is
becoming an active frontier to explore both theoretically and
experimentally.

\section*{Acknowledgements}
I would like to express my heartful acknowledgements for the
discussions and collaborations with a number of my students: Zhi
Xiao, Shi-Min Yang, Lijing Shao, Lingli Zhou, Xinyu Zhang, Yunqi Xu,
and Nan Qin, who devoted their wisdoms and enthusiasms bravely on
the topic of Lorentz violation in the past a few years. The content
of this report is mainly based on the collaborated works with them.
I am very indebted to Prof.Chao-Qiang Geng for his suggestion for
the title of this article and to many colleagues for the helpful
discussions during several workshops and seminars. The work was
supported by National Natural Science Foundation of China
(Nos.~10975003, 11021092, 11035003 and 11120101004).


\end{document}